\documentclass[twocolumn,A4]{article}
\usepackage[dvips]{graphics}
\usepackage{setspace}
\usepackage{color}
\definecolor{gold}{rgb}{0.85,0.66,0}
\definecolor{dblue}{rgb}{0,0,0.8}
\begin{document}
\onecolumn
\begin{center}
{\bf{\Large {\textcolor{gold}{NOR gate response in a double quantum ring: 
An exact result}}}}\\
~\\
{\textcolor{dblue}{Santanu K. Maiti}}$^{1,2,*}$ \\
~\\
{\em $^1$Theoretical Condensed Matter Physics Division,
Saha Institute of Nuclear Physics, \\
1/AF, Bidhannagar, Kolkata-700 064, India \\
$^2$Department of Physics, Narasinha Dutt College,
129, Belilious Road, Howrah-711 101, India} \\
~\\
{\bf Abstract}
\end{center}
NOR gate response in a double quantum ring, where each ring is threaded by a 
magnetic flux $\phi$, is investigated. The double quantum ring is sandwiched
symmetrically between two semi-infinite one-dimensional metallic electrodes 
and two gate voltages, namely, $V_a$ and $V_b$, are applied, respectively, 
in lower arms of the two rings those are treated as the two inputs of the 
NOR gate. A simple tight-binding model is used to describe the system and 
all the calculations are done through the Green's function formalism. Here 
we exactly calculate the conductance-energy and current-voltage 
characteristics as functions of the ring-to-electrode coupling strengths, 
magnetic flux and gate voltages. Our numerical study predicts that, for a 
typical value of the magnetic flux $\phi=\phi_0/2$ ($\phi_0=ch/e$, the 
elementary flux-quantum), a high output current ($1$) (in the logical 
sense) appears if both the inputs to the gate are low ($0$), while if one 
or both are high ($1$), a low output current ($0$) results. It clearly 
demonstrates the NOR gate behavior and this aspect may be utilized in 
designing an electronic logic gate. 

\vskip 0.4cm
\begin{flushleft}
\end{flushleft}
\vskip 4.4in
{\bf ~$^*$Corresponding Author}: Santanu K. Maiti

Electronic mail: santanu.maiti@saha.ac.in

\newpage
\twocolumn

\section{Introduction}

Low dimensional model quantum systems like, quantum rings, quantum dots, 
arrays of quantum rings and dots, etc, have been the objects of intense
research, both in theory and in experiments, mainly due to the fact that
these simple looking systems are prospective candidates for nano devices
in electronic as well as spintronic engineering. The key idea of 
manufacturing nano devices is based on the concept of quantum
interference effect, and it is generally preserved in the samples 
of much smaller sizes. While, the effect disappears for larger systems. 
A mesoscopic metallic ring is a promising example where electronic 
motion is confined, and for small enough in size, electron transport 
becomes phase-coherent throughout the ring~\cite{webb}. With the aid 
of two such metallic rings, we construct a double quantum ring, and, 
here we will 
explore how such a simple geometric model can be used in designing 
an electronic NOR gate. To reveal this fact, we make a 
bridge system by inserting the double quantum ring between two 
external electrodes, namely, source and drain (see Fig.~\ref{nor}).
This is the so-called source-double quantum ring-drain bridge. 
The theoretical description of electron transfer in a bridge system
has got much progress following the pioneering work of Aviram and
Ratner~\cite{aviram}. Later, several excellent 
experiments~\cite{tali,reed1,reed2} have been done in different bridge
systems to understand the basic mechanisms underlying the electron
transport. Though in literature many theoretical~\cite{orella1,orella2,
nitzan1,nitzan2,new,muj1,muj2,walc2,walc3,cui,baer2,baer3,tagami,walc1,
baer1} as well as experimental works~\cite{tali,reed1,reed2} on electron
transport are available, yet lot of controversies are still present
between the theory and experiment, and the complete knowledge of the
conduction mechanism in this scale is not very well established even today.

The main focus of the present work is to describe the NOR gate response in 
a double quantum ring where each ring is threaded by a magnetic flux $\phi$. 
The double quantum ring is sandwiched symmetrically between the electrodes, 
and the lower arms of the ring are subjected to the gate voltages $V_a$ and 
$V_b$, respectively, those are considered as the two inputs of the NOR gate 
(see Fig.~\ref{nor}). Here we use a simple tight-binding model to describe 
the system and we perform all the calculations numerically. The NOR gate 
behavior is addressed by studying the conductance-energy and current-voltage 
characteristics in terms of the ring-to-electrode coupling strengths, 
magnetic flux and gate voltages. Our numerical results propose that 
for a typical value of the magnetic flux, $\phi=\phi_0/2$, a high 
output current ($1$) (in the logical sense) is available only when both 
the two inputs
to the gate are low ($0$). While if anyone or both are high ($1$), a low 
output current ($0$) results. This feature clearly demonstrates the NOR
gate behavior and it may be utilized in manufacturing an electronic logic 
gate. To the best of our knowledge the NOR gate response in such a simple 
system has not been described earlier in the literature.

The paper is arranged as follow. Following the introduction (Section $1$), 
in Section $2$, we present the model and the theoretical formulations for 
our calculations. Section $3$ discusses the significant results, and 
finally, we summarize our results in Section $4$.

\section{Model and the synopsis of the theoretical background}

\begin{figure}[ht]
{\centering \resizebox*{7.5cm}{6.5cm}{\includegraphics{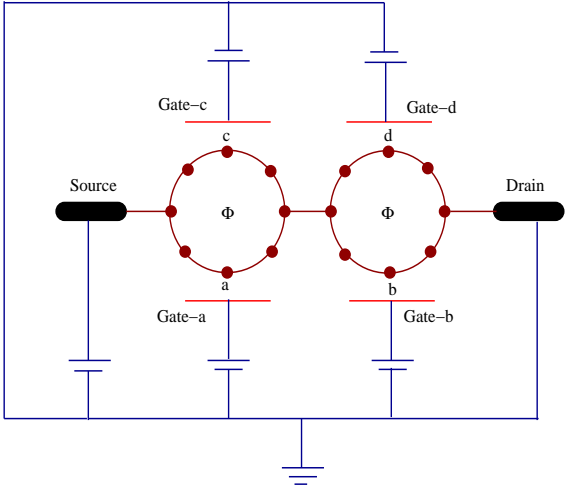}}\par}
\caption{(Color online). The scheme of connections with the batteries
for the operation of the NOR gate.}
\label{nor}
\end{figure}

We begin by referring to Fig.~\ref{nor}. A double quantum ring is 
sandwiched symmetrically between two semi-infinite one-dimensional 
($1$D) metallic electrodes. Each ring is threaded by a magnetic flux
$\phi$, the so-called Aharonov-Bohm (AB) flux which is the key 
controlling parameter for the whole operation of the NOR gate. The 
atomic sites $a$ and $b$ in the lower arms of the two rings are 
subjected to the gate voltages $V_a$ and $V_b$ through the gate electrodes
gate-a and gate-b, respectively. These gate voltages are variable and 
treated as the two inputs of the NOR gate. In a similar way we also apply 
two other gate voltages $V_c$ and $V_d$, those are not varying, in the atomic
sites $c$ and $d$ in the upper arms of the two rings via the gate electrodes
gate-c and gate-d, respectively. All these gate electrodes are ideally
isolated from the rings, and here we assume that the gate voltages 
each operate on the atomic sites nearest to the plates only. While, in 
complicated geometric models, the effect must be taken into account for 
the other atomic sites, though the effect becomes too small. The actual 
scheme of connections with the batteries for the operation of the NOR 
gate is clearly presented in the figure (Fig.~\ref{nor}), where the 
source and the gate voltages are applied with respect to the drain.

Based on the Landauer conductance formula~\cite{datta,marc} we determine the
conductance ($g$) of the double quantum ring. At much low temperatures and 
bias voltage it can be expressed in the form,
\begin{equation}
g=\frac{2e^2}{h} T
\label{equ1}
\end{equation}
where $T$ gives the transmission probability of an electron across the 
double quantum ring. In terms of the Green's function of the double
quantum ring and its coupling to the electrodes, the transmission 
probability can be represented by the relation~\cite{datta,marc},
\begin{equation}
T={\mbox{Tr}} \left[\Gamma_S G_{R}^r \Gamma_D G_{R}^a\right]
\label{equ2}
\end{equation}
where $G_{R}^r$ and $G_{R}^a$ are respectively the retarded and advanced
Green's functions of the double quantum ring including the effects of 
the electrodes. The parameters $\Gamma_S$ and $\Gamma_D$ describe the 
coupling of the double quantum ring to the source and drain, respectively. 
For the complete system i.e., the double quantum ring, source and drain, 
the Green's function is defined as,
\begin{equation}
G=\left(E-H\right)^{-1}
\label{equ3}
\end{equation}
where $E$ is the injecting energy of the source electron. In order to 
evaluate this Green's function, the inversion of an infinite matrix is 
required since the complete system consists of the finite size double
quantum ring and the two semi-infinite metallic
electrodes. However, the entire system can be divided into sub-matrices 
corresponding to the individual sub-systems and the effective Green's 
function for the double quantum ring can be written as,
\begin{equation}
G_{R}=\left(E-H_{R}-\Sigma_S-\Sigma_D\right)^{-1}
\label{equ4}
\end{equation}
where $H_{R}$ is the Hamiltonian of the double quantum ring. Withing the
approximation of the non-interacting picture, the Hamiltonian can be
expressed as,
\begin{eqnarray}
H_{R} & = & \sum_i \left(\epsilon_i + V_a \delta_{ia} + V_b \delta_{ib}
+ V_c \delta_{ic} + V_d \delta_{id} \right) \nonumber \\
 &  & c_i^{\dagger} c_i + \sum_{<ij>} 
t \left(c_i^{\dagger} c_j e^{i\theta}+ c_j^{\dagger} c_i e^{-i\theta}\right)
\label{equ5}
\end{eqnarray}
In this Hamiltonian, $\epsilon_i$'s are the site energies for all the 
sites $i$, except the sites $i=a$, $b$, $c$ and $d$ where the gate voltages 
$V_a$, $V_b$, $V_c$ and $V_d$ are applied. These gate voltages can be 
incorporated through the site energies as expressed in the above 
Hamiltonian. $c_i^{\dagger}$ ($c_i$) is the creation (annihilation) 
operator of an electron at the site $i$ and $t$ is the hopping integral
among the two neighboring sites in each ring. The hopping strength between
the two sites through which the rings are coupled to each other is also set
to $t$, for the sake of simplicity. The phase factor $\theta$ follows the
relation, $\theta=2 \pi \phi/N \phi_0$, which comes due to the penetration
of the AB flux $\phi$ through the ring, where $N$ represents the total 
number of atomic sites (filled red circles) in each ring. The effect of 
the magnetic flux $\phi$ enters explicitly into the above Hamiltonian 
(Eq.~(\ref{equ5})), and since no magnetic field is penetrated anywhere 
in the circumferences of the two rings, the above Hamiltonian is free from 
any Zeeman term. A similar kind of tight-binding Hamiltonian is also 
taken into account, except the phase factor $\theta$, to describe the 
semi-infinite $1$D metallic electrodes where the Hamiltonian 
is parametrized by constant on-site potential $\epsilon^{\prime}$ and 
nearest-neighbor hopping integral $t^{\prime}$. The double quantum ring 
is coupled to the electrodes by the parameters $\tau_S$ and $\tau_D$, 
those correspond to the coupling strengths to the source and drain, 
respectively, and they (coupling parameters) enter into the terms 
$\Sigma_S$ and $\Sigma_D$ (see Eq.~(\ref{equ4}))~\cite{datta}. These
factors ($\Sigma_S$ and $\Sigma_D$) represent the self-energies due 
to the coupling of the double quantum ring to the source and drain, 
respectively, where all the information of the coupling are included 
into these two self-energies. The detailed description is available
in the reference~\cite{datta}.

The current $I$ passing through the double quantum ring can be expressed
in terms of the applied bias voltage $V$ by the relation~\cite{datta},
\begin{equation}
I(V)=\frac{e}{\pi \hbar}\int \limits_{E_F-eV/2}^{E_F+eV/2} T(E)~ dE
\label{equ8}
\end{equation}
where $E_F$ is the equilibrium Fermi energy. Here we make a realistic
assumption that the entire voltage is dropped across the ring-electrode
interfaces, and it is examined that under such an assumption the $I$-$V$
characteristics do not change their qualitative features. 

In this work, all the results are computed only at absolute zero
temperature. These results are also valid even for some finite (low)
temperatures, since the broadening of the energy levels of the double
quantum ring due to its coupling to the electrodes becomes much larger
than that of the thermal broadening~\cite{datta}. On the other hand,
at high temperature limit, all these features completely disappear.
This is due to the fact that the phase coherent length decreases
significantly with the rise of temperature where the contribution
comes mainly from the scattering on phonons, and therefore, the
quantum interference effect vanishes. For the sake of simplicity,
we take the unit $c=e=h=1$ in our present study.

\section{Results and discussion}

Let us start our discussion by mentioning the values of the different
parameters associated with the numerical calculations. In the double
\begin{figure}[ht]
{\centering \resizebox*{8cm}{7cm}{\includegraphics{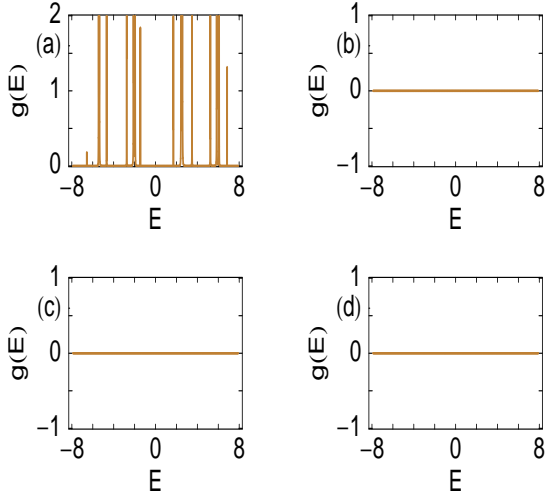}}\par}
\caption{(Color online). $g$-$E$ spectra for a double quantum ring with
$M=16$, $V_c=V_d=2$ and $\phi=0.5$ in the limit of weak-coupling.
(a) $V_a=V_b=0$, (b) $V_a=2$ and $V_b=0$, (c) $V_a=0$ and $V_b=2$ and
(d) $V_a=V_b=2$.}
\label{condlow}
\end{figure}
quantum ring, the on-site energy $\epsilon_i$ is set at $0$ for all
the atomic sites $i$, except the sites $a$, $b$, $c$ and $d$, where 
the site energies are considered as $V_a$, $V_b$, $V_c$ and $V_d$,
respectively, and the nearest-neighbor hopping strength $t$ is fixed 
at $3$. On the other hand, for the two side attached $1$D metallic
electrodes the on-site energy ($\epsilon^{\prime}$) and the 
nearest-neighbor hopping integral ($t^{\prime}$) are chosen as $0$ and 
$4$, respectively. The gate voltages $V_c$ and $V_d$, those are not
varying, are fixed at the value $2$ i.e., $V_c=V_d=2$, and the Fermi 
energy $E_F$ is set at $0$. Throughout the discussion, we focus all 
the essential features of electron transport 
for the two limiting cases depending on the strength of the coupling 
of the double quantum ring to the source and drain. Case $I$: The 
weak-coupling limit. It is specified by the condition $\tau_{S(D)} << t$. 
For this regime we choose $\tau_S=\tau_D=0.5$. Case $II$: The 
strong-coupling limit. This is mentioned by the condition $\tau_{S(D)} 
\sim t$. In this particular regime, we set the values of the parameters 
\begin{figure}[ht]
{\centering \resizebox*{8cm}{7cm}{\includegraphics{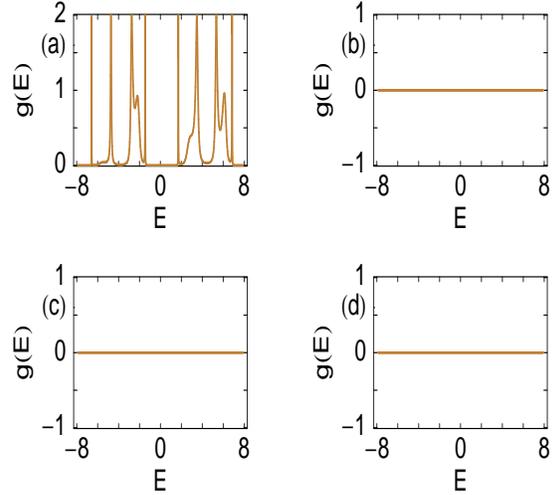}}\par}
\caption{(Color online). $g$-$E$ spectra for a double quantum ring with
$M=16$, $V_c=V_d=2$ and $\phi=0.5$ in the limit of strong-coupling.
(a) $V_a=V_b=0$, (b) $V_a=2$ and $V_b=0$, (c) $V_a=0$ and $V_b=2$ and
(d) $V_a=V_b=2$.}
\label{condhigh}
\end{figure}
as $\tau_S=\tau_D=2.5$. The key controlling parameter for all these 
calculations is the AB flux $\phi$, threaded by each ring, which 
is fixed at $\phi_0/2$ i.e., $0.5$ in our chosen unit $c=e=h=1$.

As representative examples, in Fig.~\ref{condlow} we display the variation
of the conductance $g$ as a function of the injecting electron energy $E$
for a double quantum ring with $M=16$ ($M=2N$, the total number of atomic 
sites in the double quantum ring, since each ring contains $N$ atomic 
sites) in the weak-coupling limit, where
(a), (b), (c) and (d) represent the results for the four different cases 
of the gate voltages $V_a$ and $V_b$, respectively. Quite interestingly 
from these spectra we observe that, for the case when both the two inputs 
$V_a$ and $V_b$ are identical to $2$ i.e., both are high, the conductance 
$g$ becomes exactly zero for the full range of the energy $E$ (see 
Fig.~\ref{condlow}(d)). The exactly similar response is also visible for
the two other cases where anyone of the two inputs is high and other is
low. The results are shown in Figs.~\ref{condlow}(b) and (c), respectively.
Hence, for all these three cases (Figs.~\ref{condlow}(b)-(d)), no electron
conduction takes place from the source to the drain through the double
quantum ring. The electron conduction through the bridge system is allowed
only for the typical case where both the two inputs two the gates are low
i.e., $V_a=V_b=0$. The spectrum is given in Fig.~\ref{condlow}(a). It is
noticed that, for some particular energies the conductance exhibits sharp
resonant peaks. At these resonant energies, the conductance approaches 
\begin{figure}[ht]
{\centering \resizebox*{8cm}{7cm}{\includegraphics{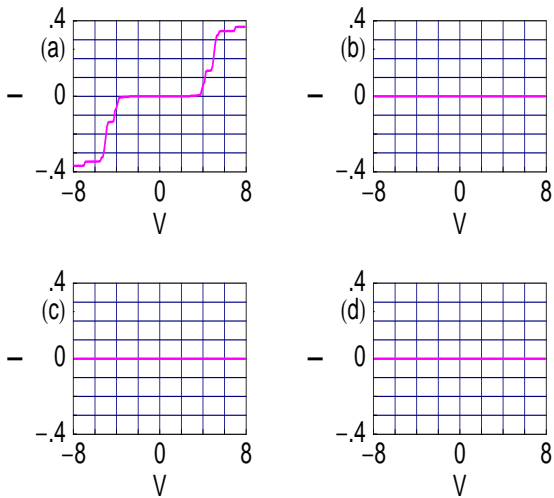}}\par}
\caption{(Color online). Current $I$ as a function of the applied bias
voltage $V$ for a double quantum ring with $M=16$, $V_c=V_d=2$ and
$\phi=0.5$ in the weak-coupling limit. (a) $V_a=V_b=0$, (b) $V_a=2$ and
$V_b=0$, (c) $V_a=0$ and $V_b=2$ and (d) $V_a=V_b=2$.}
\label{currlow}
\end{figure}
the value $2$, and accordingly, the transmission probability $T$ goes 
to unity, since the relation $g=2T$ holds from the Landauer conductance 
formula (see Eq.~(\ref{equ1}) with $e=h=1$). All these resonant peaks 
are associated with the energy eigenvalues of the double quantum ring, 
and therefore, it can be emphasized that the conductance spectrum 
manifests itself the electronic structure of the double quantum ring.
Now we try to explain the roles of the gate voltages on the electron 
transport in these four different cases. The probability amplitude of 
getting an electron from the source to drain across the double quantum 
ring depends on the combined effect of the quantum interferences of the 
electronic waves passing through the upper and lower arms of the two rings.
For a symmetrically connected ring (lengths of the two arms of the ring 
are identical to each other), threaded by an AB flux $\phi$, 
the probability amplitude of getting an electron across the ring becomes 
exactly zero ($T=0$) for the typical flux, $\phi=\phi_0/2$. This is due 
to the result of the quantum interference among the two waves in the two 
arms of the ring~\cite{webb}, which can be shown through few simple 
mathematical steps.
Thus for the particular case when both the two inputs to the gate are high 
i.e., $V_a=V_b=2$, the upper and lower arms of the two rings become exactly 
identical since the gate voltages $V_c$ and $V_d$ in the upper arms are also
fixed at the value $2$. This provides the vanishing transmission probability. 
If the input voltages $V_a$ and $V_b$ are different from the potential 
applied in the atomic sites $c$ and $d$, then the upper and the lower arms 
of the two rings are no longer identical to each other and the transmission 
probability will not vanish. Thus, to get the zero transmission probability 
\begin{table}[ht]
\begin{center}
\caption{NOR gate response in the limit of weak-coupling. The current
$I$ is computed at the bias voltage $6.02$.}
\label{table1}
~\\
\begin{tabular}{|c|c|c|}
\hline \hline
Input-I ($V_a$) & Input-II ($V_b$) & Current ($I$) \\ \hline
$0$ & $0$ & $0.346$ \\ \hline
$2$ & $0$ & $0$ \\ \hline
$0$ & $2$ & $0$ \\ \hline
$2$ & $2$ & $0$ \\ \hline \hline
\end{tabular}
\end{center}
\end{table}
when the inputs are high, we should tune $V_c$ and $V_d$ properly, observing 
the input potentials and vice versa. The similar behavior is also noticed 
for the two other cases ($V_a=2$, $V_b=0$ and 
$V_a=0$, $V_b=2$), where the symmetry is broken in only one ring out of 
these two by making the gate voltage either in the site $b$ or in $a$ 
to zero, maintaining the symmetry in the other ring. The reason is that, 
when anyone of the two gates ($V_a$ and $V_b$) becomes zero, the symmetry 
between the upper and lower arms is broken only in one ring which provides 
non-zero transmission probability across the ring. While, for the other ring 
where the gate voltage is applied, the symmetry between the two arms becomes 
preserved which gives zero transmission probability. Accordingly, the 
combined effect of these two rings provides vanishing transmission 
probability across the 
bridge, as the two rings are coupled to each other. The non-zero value of 
the transmission probability appears only when the symmetries of both 
the two rings are identically broken, and it is available for the particular
case when both the two inputs to the gate are low i.e., $V_a=V_b=0$. This 
feature clearly demonstrates the NOR gate behavior.
With these properties, we get additional one feature when the coupling 
strength of the double quantum ring to the electrodes is increased from 
the low regime to the high one. In the limit of strong-coupling, all 
these resonant peaks get substantial widths compared to the weak-coupling 
\begin{figure}[ht]
{\centering \resizebox*{8cm}{7cm}{\includegraphics{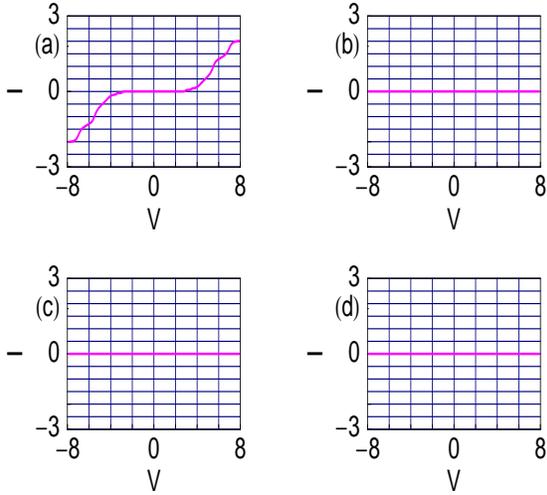}}\par}
\caption{(Color online). Current $I$ as a function of the applied bias
voltage $V$ for a double quantum ring with $M=16$, $V_c=V_d=2$ and
$\phi=0.5$ in the strong-coupling limit. (a) $V_a=V_b=0$,
(b) $V_a=2$ and $V_b=0$, (c) $V_a=0$ and $V_b=2$ and (d) $V_a=V_b=2$.}
\label{currhigh}
\end{figure}
limit. The results are shown in Fig.~\ref{condhigh}, where all the other 
parameters are identical to those in Fig.~\ref{condlow}. The contribution 
for the broadening of the resonant peaks in this strong-coupling limit 
appears from the imaginary parts of the self-energies $\Sigma_S$ and 
$\Sigma_D$, respectively~\cite{datta}. Hence, by tuning the coupling 
strength, we can get the electron transmission across the double quantum 
ring for the wider range of energies and it provides an important signature 
in the study of the current-voltage ($I$-$V$) characteristics.

All these features of electron transfer become much more clearly visible
by studying the $I$-$V$ characteristics. The current passing through the
double quantum ring is computed from the integration procedure of the
transmission
function $T$ as prescribed in Eq.~(\ref{equ8}). The transmission
function varies exactly similar to that of the conductance spectrum,
differ only in magnitude by the factor $2$ since the relation $g=2T$
holds from the Landauer conductance formula Eq.~(\ref{equ1}).
As illustrative examples, in Fig.~\ref{currlow} we show the variation of
the current $I$ as a function of the applied bias voltage $V$ for a 
double quantum ring with $M=16$ in the limit of weak-coupling, 
where (a), (b), (c) and (d) correspond to the results for the 
different cases of the two input voltages, respectively.
For the cases when either both the two inputs to the gate are high 
($V_a=V_b=2$), or anyone of the two inputs is high and other is low 
($V_a=2$, $V_b=0$ or $V_a=0$, $V_b=2$), the current drops exactly to zero 
for the whole range of the bias voltage. The results are shown in 
Figs.~\ref{currlow}(b)-(d), and, the vanishing behavior of the current 
in these three different cases can be clearly understood from the 
conductance spectra given in Figs.~\ref{condlow}(b)-(d), since the
current is computed from the integration procedure of the transmission
function $T$.
\begin{table}[ht]
\begin{center}
\caption{NOR gate response in the limit of strong-coupling. The current
$I$ is computed at the bias voltage $6.02$.}
\label{table2}
~\\
\begin{tabular}{|c|c|c|}
\hline \hline
Input-I ($V_a$) & Input-II ($V_b$) & Current ($I$) \\ \hline
$0$ & $0$ & $1.295$ \\ \hline
$2$ & $0$ & $0$ \\ \hline
$0$ & $2$ & $0$ \\ \hline
$2$ & $2$ & $0$ \\ \hline \hline
\end{tabular}
\end{center}
\end{table}
The finite value of the current is observed only for the typical case 
where both the two inputs to the gate are low i.e., $V_a=V_b=0$. The 
result is shown in Fig.~\ref{currlow}(a). From this figure it is observed 
that the current exhibits staircase-like structure with fine steps as a 
function of the applied bias voltage. This is due to the existence of the 
sharp resonant peaks in the conductance spectrum in the weak-coupling limit, 
since the current is computed by the integration method of the transmission 
function $T$. With the increase of the bias voltage $V$, the electrochemical 
potentials on the electrodes are shifted gradually, and finally cross one 
of the quantized energy levels of the double quantum ring. Therefore, a 
current channel is opened up which produces a jump in the $I$-$V$ 
spectrum. Addition to these behaviors, it is also important to note that 
the non-zero value of the current appears beyond a finite value of $V$, 
the so-called threshold voltage ($V_{th}$). This $V_{th}$ can be tuned 
by controlling the size ($N$) of the two rings. From these $I$-$V$ 
spectra, the behavior of the NOR gate response is clearly observed. To make 
it more clearer, in Table~\ref{table1}, we present a quantitative estimate 
of the typical current amplitude, computed at the bias voltage $V=6.02$, 
in this weak-coupling limit. It shows $I=0.346$ only when both the two 
inputs to the gate are low ($V_a=V_b=0$), while for the other three cases 
when either $V_a=V_b=2$ or $V_a=2$, $V_b=0$ or $V_a=0$, $V_b=2$, the current
$I$ gets the value $0$.
In the same footing, as above, here we also discuss the $I$-$V$
characteristics in the limit of strong-coupling. In this limit, the
current varies almost continuously with the applied bias voltage and
achieves much larger amplitude than the weak-coupling case
(Fig.~\ref{currlow}), as presented in Fig.~\ref{currhigh}. The reason
is that, in the limit of strong-coupling all the resonant peaks get
broadened which provide larger current in the integration procedure
of the transmission function $T$. Thus by tuning the strength of the
ring-to-electrodes coupling, we can achieve very large current amplitude
from the very low one for the same bias voltage $V$. All the other
properties i.e., the dependences of the gate voltages on the $I$-$V$
characteristics are exactly similar to those as given in Fig.~\ref{currlow}.
In this strong-coupling limit we also make a quantitative study for the
typical current amplitude, given in Table~\ref{table2}, where the
current amplitude is determined at the same bias voltage ($V=6.02$) as
earlier. The response of the output current is exactly similar to that as
given in Table~\ref{table1}. Here the non-zero value of the current gets
the value $1.295$ which is much larger compared to the weak-coupling case
that provides the value $0.346$. From these results the NOR gate response
in a double quantum ring can be clearly manifested.

\section{Concluding remarks}

In conclusion, we have explored the NOR gate response in a double quantum
ring where each ring is threaded by a magnetic flux $\phi$. The double
quantum ring is attached symmetrically to two semi-infinite $1$D metallic
electrodes, and two gate voltages, namely, $V_a$ and $V_b$, are applied,
respectively, in the lower arms of the two rings and they are treated as
the two inputs of the NOR gate. The model is described by the 
tight-binding Hamiltonian, and all the calculations are performed within
the Green's function formalism. We have done exact numerical calculation
to determine the conductance-energy 
and current-voltage characteristics as functions of the ring-electrode 
coupling strengths, magnetic flux and gate voltages. Very nicely we have 
noticed that, for the half flux-quantum value of $\phi$ ($\phi=\phi_0/2$), 
a high output current ($1$) (in the logical sense) appears if both the 
two inputs to the gate are low ($0$). While, if one or both are 
high ($1$), a low output current ($0$) results. It clearly manifests 
the NOR gate response and this aspect may be utilized in designing a 
tailor made electronic logic gate. 

Throughout our presentation, we have addressed the conductance-energy 
and current-voltage characteristics for a double quantum ring with total 
number of atomic sites $M=16$. In our model calculations, this typical 
number ($M=16$) is chosen only for the sake of simplicity. Though the 
results presented here change numerically with the ring size ($N$), but 
all the basic features remain exactly invariant. To be more specific, it
is important to note that, in real situation the experimentally 
achievable rings have typical diameters within the range $0.4$-$0.6$ $\mu$m. 

In the present work we have done all the calculations by ignoring
the effects of the temperature, electron-electron correlation, disorder,
etc. Due to these factors, any scattering process that appears in the
arms of the rings would have influence on electronic phases, and, in
consequences can disturb the quantum interference effects. Here we
have assumed that in our sample all these effects are too small, and
accordingly, we have neglected all these factors in this particular
study.

The importance of this article is mainly concerned with (i) the simplicity 
of the geometry and (ii) the smallness of the size. To the best of our 
knowledge the NOR gate response in such a simple low-dimensional system 
that can be operated even at finite temperatures (low) has not been 
addressed earlier in the literature.

At the end, here we have designed a NOR gate using mesoscopic rings, 
based on the effect of quantum interference, which is a classical logic
gate. On the other hand, quantum logic gates using such rings have 
already been proposed earlier which can be available in the 
reference~\cite{peeters}.

\end{document}